\documentstyle[prl,aps]{revtex}
\input epsf.sty
\begin{document}
\draft
\draft
\twocolumn[\columnwidth\textwidth\csname@twocolumnfalse\endcsname
\title {Rotations in isospace: a doorway to the understanding \\
 of neutron-proton superfluidity in N=Z nuclei}
\author{Wojciech Satu{\l}a$^{(1,2)}$ and Ramon Wyss$^{(1)}$}
\address{$^{(1)}$ Royal Institute of Technology, Frescativ. 24,
S-104 05 Stockholm, Sweden}
\address{$^{(2)}$Institute of Theoretical Physics, Warsaw University,
ul. Ho\.za 69, PL-00 681, Warsaw, Poland }
\date{\today}
\maketitle
\begin{abstract}
The $T$=2 excitations in even-even $N$=$Z$ nuclei are calculated 
within the isospin cranked mean-field approach. 
The response of pairing correlations to rotation
in isospace is investigated.   
It is shown that whereas the isovector pairing rather modestly
modifies the single-particle moment of inertia in isospace, the  
isoscalar pairing strongly reduces its value. This reduction of
the moments of inertia in isospace with respect to its rigid
body value is a strong indicator of
collective isoscalar pairing correlations.
Beautiful analogies between the role of isovector pairing for the case
of spatial rotations and the role of isoscalar pairing for the case 
of iso-rotations are underlined. 
\end{abstract}
\addvspace{5mm}]
\narrowtext
The ground state of most nuclei can be characterized in terms of a
superfluid condensate of Cooper pairs which are formed by
nucleons moving in time reversed orbits~\cite{[Boh58]}. 
Nuclei with identical number of protons and neutrons, $N$=$Z$, exhibit
an additional  symmetry, related to the similarity of proton and neutron
wave functions at the Fermi surface. 
In such a case protons and neutrons occupying identical spatial 
orbitals may form
two fundamentally different Cooper pairs of either isovector ($T$=1) 
or isoscalar ($T$=0) type~\cite{[Goo79]}. 
The question whether isoscalar 
pairing may form a condensate similar to the well 
established isovector pairing
has gained considerable interest in recent time.

Already early on it was noticed that the understanding of
excitation energies of the  isobaric analogue states 
yield important information on the effective nuclear
force, in particular also on its pairing 
component~\cite{[Jan65],[Zel76]}, see also recent 
works~\cite{[Vog00],[Mac00s]}. 
Therefore, we analyze the $T$=2  excitations
in even-even $N$=$Z$ nuclei by means 
of the cranking approximation in 
isospace which is the lowest (linear) order approximation to
the projection onto good isospin~\cite{[Kam68]}. 
This approximation 
has been tested within an exactly solvable model by Chen {\it 
et al.\/}~\cite{[Che78s]} where it was concluded that,
in conjunction with number-projection, it offers a 
reliable approximation to the exact solutions.
Another motivation to apply the cranking approximation 
stems from the formal analogy between spatial and isospin
rotations following  
$a_I$$I(I+1)$ and $a_T$$T(T+1)$ patterns, respectively~\cite{[Boh69]}.
The crucial quantity of our investigation is the inertia  parameter
in isospace,  $a_{T}$
[reciprocal of the moment of inertia $\Im_{T}$].
Indeed, the study of the rotational spectra  
was crucial to establish
evidence for superfluidity in atomic nuclei~\cite{[Bel59]}.
Similarly, our microscopic calculations of the nuclear 
inertia parameter $a_T$ in isospace
show that it strongly depends
on the short-range pairing correlations.
However, in contrast to spatial rotations
it is shown that $a_T$ is rather insensitive to isovector
but extremely  sensitive to isoscalar 
proton-neutron pairing correlations.
\begin{figure}[thb]
\begin{center}
\leavevmode
\epsfysize=8.0cm
\epsfbox{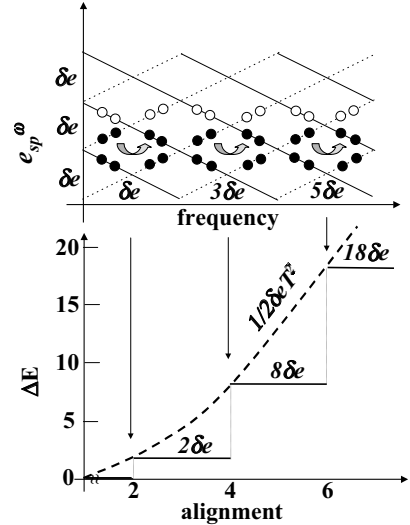}
\end{center}
  \caption{The single-particle routhians (upper panel) versus
  the iso-cranking frequency for the equidistant level model. Solid (dashed)
  lines depict isovector (isoscalar) $sp$ states, respectively.
  At each crossing frequency (indicated by arrows) the configuration changes,
  and hence excitation energy and iso-alignment (lower panel).
  }
  \label{fig1}
  \end{figure}
Before entering the details of our model, let us consider
a single-particle ($sp$) routhian $\hat H^\omega = \hat H_{sp} -
\omega \hat t_x$. For simplicity, let us also assume
that the spectrum of $\hat H_{sp}$ is equidistant
[$e_i=i\delta e$]  and iso-symmetric. Hence, at $\omega =0$ each
eigenstate of $\hat H^\omega$ is
four-fold degenerate.
The cranking term, $-\omega \hat t_x$, lifts the isospin but not 
Kramers degeneracy resulting in isovector
$|iv\rangle=\frac{1}{\sqrt2}(|n\rangle - |p\rangle )$
and isoscalar $|is\rangle =\frac{1}{\sqrt2}
(|n\rangle +|p\rangle)$ doublets, with isospin alignment 
of $\mp 1/2$, respectively (Fig.~1). Clearly, the
ground state configuration changes stepwise at the crossing frequencies:
$\omega_{c}^{(n)} = \delta e, 3\delta e, 5\delta e, \cdots ,(2n-1)\delta e$.
At each crossing frequency two isoscalar states are emptied and
two isovector become  occupied. Hence, the total isospin alignment,
$\langle \hat t_x \rangle\equiv T_x$, changes in steps
of $\Delta T_x =2$.
Since at the same time $T_z=T_y\equiv 0$,
i.e. $\Delta T_x \equiv \Delta T$, the stepwise
increase corresponds to {\it non-collective rotation\/}.
The ground state band (gsb) consists of only even isospin states,
$T = 0,2,4,\cdots ,2n$ in complete analogy to the even
spin sequence in the gsb of spatially rotating even-even nuclei.
Exploring further
this analogy, one can show that the even--$T$ sequence in the gsb
is a consequence of iso-signature
symmetry, $\hat R_\tau =\exp (-i\pi\hat t_x)$, similar to
signature symmetry conservation, $\hat R =\exp (-i\pi\hat j_x)$,  for
rotational motion. It is of importance to underline that
{\sl odd--$T$ states
can be reached only by the proper particle-hole excitation at
$\omega =0$}.

Once the crossing frequencies are calculated, it is straightforward
to compute the excitation energy $E_T$ (with respect to the
$gs$) spectrum of the iso-rotational gsb band:
\begin{equation}\label{gsb}
E_T = E^\omega +\omega T_x = 2\displaystyle\sum_{i=1}^{T_x/2}\omega_c^{(i)}
 =  {1\over 2} \delta e T_x^2.
\end{equation}
The schematic $sp$ model leads to the classical rotational 
formula $\sim$$T_x^2$ with an inertia parameter proportional
to the single-particle splitting at the Fermi energy.


  \begin{figure}[htb]
\begin{center}
\leavevmode
\epsfysize=9.0cm
\epsfbox{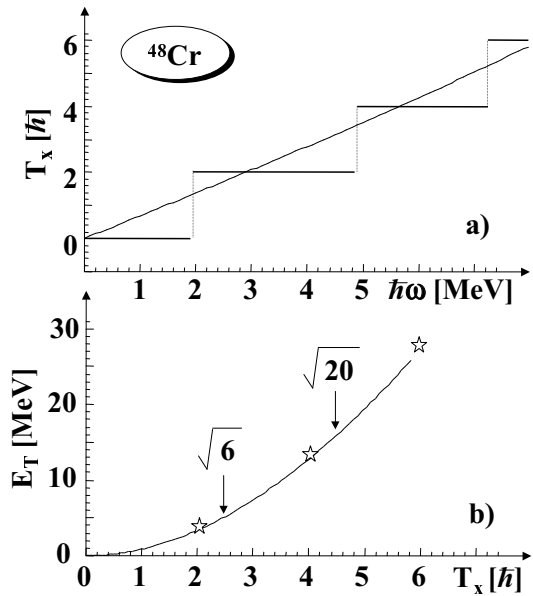}
\end{center}
  \caption{Alignment versus frequency {\bf (a)} and excitation
  energy versus alignment {\bf (b)} for the $sp$ model
  (discrete step-function and $\star$) and for a model including isovector
  pairing correlations. Calculations have been performed for
  $^{48}$Cr at fixed deformation $\beta_2 = 0.25$.
  }
  \label{fig2}
  \end{figure}

Let us now briefly investigate how the $sp$ model
is affected by the presence of isovector pairing
correlations. To study this issue we have performed
a series of Lipkin-Nogami calculations for selected
$N$=$Z$ nuclei using the deformed Woods-Saxon potential
as a mean-field (at fixed deformation)
and standard isovector seniority-type pairing interaction~\cite{[Sat97a]}.
A representative example  reflecting
the generic results of the study is illustrated in Fig.~2.
The major modification introduced by
isovector pairing correlations is
the smooth increase of  iso-alignment with cranking frequency,
see Fig.~2a. The isovector pairing
introduces a kind of {\it collectivity\/}
on the top of the $sp$ model but does not affect the 
bulk properties. The value of the inertia parameter
depends on the shell-structure at the Fermi energy but in a modest
and intuitively understandable way. Namely, 
as compared to the $sp$ estimate, it
decreases (increases) 
when shell gaps are present (absent).      
It is also interesting to note that the smoothing
effect depend only weakly  on the
strength of isovector pairing correlations, $G_{T=1}$.
The alignment, as well as the
excitation energy almost do not change, even when 
the isovector pairing strength, $G_{T=1}$, is strongly
reduced.
Henceforth, the excitation energy of the $T$-states
will be computed at the standard cranking constraint
$T_x =\sqrt{T(T+1)}$.


 The main objective of our study is to 
clarify the role played by isoscalar pairing correlations, 
in particular we show that one
simultaneously: ({\it i\/})  can recover
the Wigner energy as it was shown in our previous
work~\cite{[Sat97a],[Sat00]} and ({\it ii\/}) determine the
excitation energies of the $T$=2 states in even-even $N$=$Z$ nuclei.
In addition, our subsequent publication will 
show that the mean-field
model incorporating isoscalar pairing correlations
also is able to explain on the same footing
the excitation energies of
$T$=1 states in even-even $N$=$Z$ nuclei as well as the competition of
$T$=0 and $T$=1 states in odd-odd $N$=$Z$ nuclei.
Our Hamiltonian is based on the  deformed mean-field
potential of Woods-Saxon (WS) type~\cite{[Cwi87s]}.
The two body residual interaction contains both
isovector and isoscalar seniority pairing:
\begin{equation}
\hat H^\omega = \hat h_{WS} +G_{T=1}\hat P_{1}^\dagger
\hat P_{1} +
G_{T=0} \hat P_{0}^\dagger \hat P_{0}
-  \omega \hat t_x
\label{ham}
\end{equation}
where $\hat P_{1}^\dagger$ and $\hat P_{0}^\dagger$
create isovector and isoscalar pairs, respectively.
The Hamiltonian (\ref{ham}) is solved using the Lipkin-Nogami
method. The model is very similar to the one described 
in detail in Ref.~\cite{[Sat00]}. However, different to
Ref.~\cite{[Sat00]} we now employ
the most general Bogoliubov transformation. It allows us to
fully explore the isoscalar pairing channel without
any symmetry induced restrictions
i.e. to include
simultaneously {\boldmath{$\alpha\alpha$}} and
{\boldmath{$\alpha\tilde\alpha$}}
isoscalar pairs.
In the present study, where we confine to $I$=0 states in even-even
nuclei, it is sufficient to consider 
{\boldmath{$\alpha\tilde\alpha$}}
 $T$=0 correlations.
Moreover, since this study aims at a qualitative
description, we have assumed near
spherical deformation, $\beta_2 =0.05$, for all nuclei.
A drawback of our model is the lacking response of the
isovector particle-hole (p-h) field to rotations in isospace.
To fully investigate in a quantitative way  the interplay
of isoscalar pairing and rotations in isospace,
requires self-consistent
Hartree-Fock-Bogoliubov calculations with realistic isovector
and isoscalar two-body
interactions in both p-h and pairing channels. 
This, however, is clearly beyond our present approach.
  \begin{figure}[t]
\begin{center}
\leavevmode
\epsfysize=9.0cm
\epsfbox{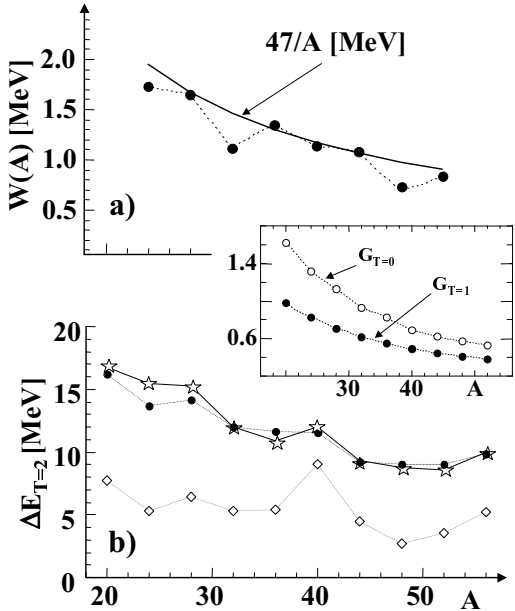}
\end{center}
  \caption{Calculated Wigner energy strength $W(A)$ {\bf (a)}
  in $N$=$Z$ nuclei.
  The insert
  shows the isovector and isoscalar pairing strength parameters used to
  reproduce the smooth empirical trend of the Wigner energy strength
  $47/A$\,MeV.  The lower panel {\bf (b)} shows the excitation energy of 
  the $T$=2, $I$=0 states. The experimental data values
  are marked by
  ($\star$); the calculations including both $T$=0
  and $T$=1 pairing are marked with
  ($\bullet$) and the calculations with $T$=1 pairing
  only are labelled with ($\diamond$).
  }
  \label{fig3}
  \end{figure}

The isovector pairing strength, $G_{T=1}$, is computed using the
average gap method of Ref.~\cite{[Mol92]} where
the number of proton (and neutron) WS states retained for the pairing
calculations is consistently put to $A$/2. To compute the strength of
the isoscalar pairing correlations, $G_{T=0}$,  we follow the
prescription given in Ref.~\cite{[Sat00]}. This method is
based on the assumption that, within the mean-field model,
the Wigner energy is predominantly due to the $T$=0 pairing
correlations. In other words we fit $G_{T=0}$ to
reproduce roughly the Wigner energy strength
$W(A)$$\approx$47/$A$\,MeV using the technique provided
in Ref.~\cite{[Sat97bs]}. The result of these calculations
is shown in Fig.~3a. Interestingly, to obtain the proper value of $W(A)$,
the mass scaling of isovector and isoscalar pairing strengths has to be
different.
The ratio $x^{T=0}$=$G_{T=0}/G_{T=1}$
necessary to reproduce the empirical trends decreases smoothly
from $x^{T=0}$$\approx$1.65 at the beginning of the $sd$-shell
to $x^{T=0}$$\approx$1.40 in the $f_{7/2}$ subshell
as shown in the insert.
The calculated and empirical excitation energies
of the lowest $T$=2 states, $\Delta E_{T=2}$, 
for even-even $N$=$Z$,  
20$\leq$$A$$\leq $56, nuclei
are displayed in Fig.~3b.
Calculations including isoscalar
pairing correlations ($\bullet$) are in excellent agreement with the empirical
data ($\star$). In contrast, calculations including only isovector pairing
field ($\diamond$) account for roughly half of the empirical
excitation energy. It is interesting to
notice that although generally
$\Delta E_{T=2}$ decreases as a function of $A$, it clearly
rises at closed (sub-)shells, particularly for $N$=$Z$=20. 
This nicely reflects the dominant role
played by the $sp$ substructure
$\Delta E_{T=2}\propto \delta e$
[see Eq.~(\ref{gsb})] and the reduced role of
the pairing correlations
(in particular the $T$=0 pairing) at shell closure. 
Note also the pronounced smoothing effect of isoscalar pairing
on the $\Delta  E_{T=2}$ excitations at shell-closure.
Last note least, it is important to remember that the Wigner energy
and the $T$=2 excitations are calculated in a totally different manner.

  \begin{figure}[t]
\begin{center}
\leavevmode
\epsfysize=8.0cm
\epsfbox{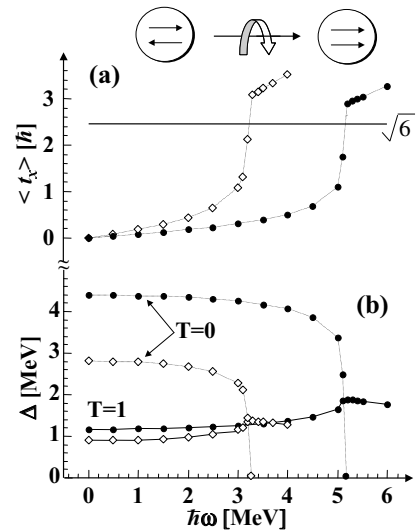}
\end{center}
  \caption{Alignment  {\bf (a)} and isoscalar
   and isovector gap parameters {\bf (b)} versus iso-cranking
   frequency calculated for $^{24}$Mg ($\bullet$) and
   $^{48}$Cr ($\diamond$). The figure illustrates the phase
   transition leading to the disappearance of isoscalar $T$=0
   pairing correlations (lower panel) once the alignment
   reaches a value of $\sqrt6$ (upper panel).
  }
  \label{fig4}
  \end{figure}

There is an appealing correspondence between the role of isoscalar
and isovector correlations. The binding energy of even-even
nuclei is lowered with respect to their odd neighbours due to the well
known blocking effect.
Analogous, the generalized blocking of
isoscalar pairing correlations lowers the
ground state of even-even $N$=$Z$ nuclei and accounts for the missing
binding energy commonly known as the Wigner energy~\cite{[Sat97a]}.
Isovector pairing is weakened at high angular momenta
due to the Coriolis effect, that tends to align the angular momenta
of the nucleons along the rotational axis.
Similarly, because the $T$=0 pairs have isospins coupled antiparallel,
rotations in  isospace tend to destroy these correlations.
In contrast, isovector pairs have their isospins coupled 
parallel and are hardly affected by iso-rotations.
Pairing correlations as a function
of rotational frequency in either space or isospace are
quenched in
a similar fashion like the magnetic-field destroys
the electronic Cooper pairs in metallic superconductors.
Hence, with increasing iso-cranking frequency one does expect
a bulk phase transition similar to the well known
Meissner effect~\cite{[Mei33]}.

The {\it phase transition\/} discussed above on general
grounds indeed occurs systematically in our calculations. 
Two representative
cases are depicted  in Fig.~\ref{fig4}.
The phase transition (cf. lower panel of Fig.~4)
takes place almost exactly at, or
just before the iso-alignment (cf. upper panel of Fig.~4)
reaches the value of
$T_x=\sqrt 6$, corresponding to the $T$=2 state.
Evidently, once we reach
the $T$=2 states, isoscalar pairing correlations
have essentially dropped to zero.
At the same time, isovector pairing correlations are still
strong. 
The quenching of isoscalar pairing in the ground state of $|N$-$Z|$=4 nuclei
is a general feature of   most known calculations, independent on 
the interaction~\cite{[Sat97a],[Sat00],[Rop00s]}.
These findings are therefore consistent with the
isobaric symmetry which demands that the structure
(and hence, also the excitation energy) of
$T$=2,$T_z$=0 states to be similar to the structure
of $T$=2,$T_z$=$\pm$2 members of the
$T$=2 quintuplet.
The $T$=2,$T_z$=$\pm$2 states are just the ground states
of the $|N$-$Z|$=4 nuclei,
since, by the rule, the ground states of even-even nuclei
are the states of minimum isospin: $T$=$|T_z|$=$|N$-$Z|$/2.

  \begin{figure}[t]
\begin{center}
\leavevmode
\epsfysize=6.0cm
\epsfbox{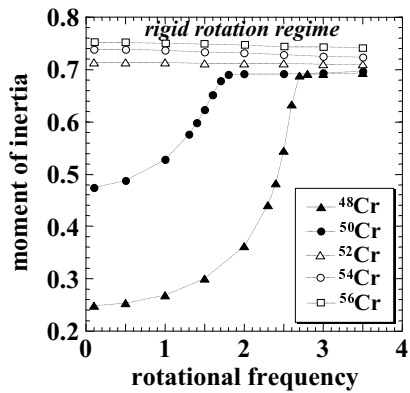}
\end{center}
  \caption{Moment of inertia (MoI), $\Im_x\equiv T_x/\omega$, 
   for a sequence of $N$-$Z$=0,2,4,6,8 Cr-isotopes at constant
   deformation ($\beta_2$=0.25)
   versus rotational frequency. The figure clearly illustrates
   the pronounced effect of the $T$=0 pair 
   correlations in $N$-$Z$=0 and 2 isotopes
   and the phase transition rising the MoI to the 'rigid-body'
   value at high frequencies. 
  }
  \label{fig5}
  \end{figure}

The dependence of the moments of inertia (MoI),
$\Im_T^{(x)}=T_x/\omega$  
as a function of 
the iso-cranking frequency
and $N$-$Z$ for a sequence of Cr-isotopes 
is further illustrated in Fig.~\ref{fig5}.
At small frequencies the $T$=0 pairing phase is present only
in the $T_z$=0,1 isotopes, where it strongly lowers the MoI.
At high frequency a {\it phase transition\/}
takes place resulting in a rapid increase of the MoI to 
the 'rigid body' value i.e. the value corresponding to 
$\Im_T^{(x)}$ for the $T$=0 unpaired system.

In summary, the response of pairing correlations to rotations
in isospace is investigated within a simple model.
The present calculations show that on a qualitative level,
the mean-field method is capable to
account for both mass-excess in $N$=$Z$ nuclei
and the MoI in isospace if and only if 
the short range correlations take into
account isoscalar pairing.
Pairing correlations of isovector and isoscalar type respond
totally different  to rotations in isospace.
The presence of isoscalar pairing strongly
reduces the MoI in isospace, but only for low values of $T$. With increasing 
iso-cranking frequency, isospin starts to align, 
iso-pairs become broken,
resulting eventually in the quenching of isoscalar
pairing and a rapid increase of the MoI. In the regime of large isospin, 
no isoscalar pairing
is present. These results are in
beautiful analogy to spatial rotations.

\bigskip
This work was supported by the G\"oran Gustafsson Foundation,
the Swedish Natural Science Research council (NFR),
and the Polish Committee for Scientific Research (KBN) under
Contract No. 2~P03B~040~14.

\end{document}